\documentclass[12pt,preprint]{emulateapj}

\shorttitle{Evolution of irradiated planets}
\shortauthors{Chabrier et al.}

\def\te{T_{\rm eff}}
\def\msol{\mbox{M}_\odot}
\def\mjup{\mbox{M}_{jup}}
\def\rjup{\mbox{R}_{jup}}

\def\tr{\mbox{OGLE-TR-56b}}
\def\hd{\mbox{HD209458b}}

\def\simgr{\,\hbox{\hbox{$ > $}\kern -0.8em \lower 1.0ex\hbox{$\sim$}}\,}
\def\simle{\,\hbox{\hbox{$ < $}\kern -0.8em \lower 1.0ex\hbox{$\sim$}}\,}

\begin{document}

\title{The evolution of irradiated planets. Application to transits}
\author{G. Chabrier$^1$, T. Barman$^2$, I. Baraffe$^1$, F. Allard$^1$, \& P.H. Hauschildt$^3$}
\affil{$^{1}$Ecole Normale Sup\'erieure de Lyon, C.R.A.L. (UMR CNRS 5574), 69364 Lyon, France\\
$^{2}$Wichita State University\\
$^{3}$Hamburger Sternwarte, Hamburg, Germany}

\begin{abstract}
Extending the theory we derived recently for $\hd$ to different cases of
strongly irradiated gaseous exoplanets, we have calculated the consistent
evolution of the new transiting planet, $\tr$, for its recently revised mass determination.
The theory is shown to successfully reproduce the observed radius, for the proper age of the system.
We also examine the dissipation of kinetic energy at the planet's internal
adiabat due to atmospheric winds, and place constraints on the efficiency of
this process.  We show that a fraction $\sim 0.1$ -- $0.5\%$ of the
incident flux transformed into thermal energy deposited at the adiabatic level
can accommodate the observed radii of both $\tr$ and $\hd$.  The present theory
yields quantitative predictions on the evolution of the emergent spectrum and
fundamental properties of hot-jupiters. The predictions for radius, luminosity,
temperature as a function of the planet's mass and orbital distance can be
used as benchmarks for future detections of transit planets.
\end{abstract}


\section{Introduction}
Since the discovery of 51 Peg b (Mayor \& Queloz 1995), over 
a hundred gaseous planets have been discovered in orbit around G, K, and M
stars.  These discoveries have opened a new domain of research in
astronomy, at the crossroad between stellar and planetary physics. Many
ground-based and space-based missions are planned for the coming years to
search for more of these objects, and the next decade will probably see the
first direct detection of an extrasolar planet. These projects require a robust
theoretical background to accurately understand the properties of these objects
and to provide reliable guidance for observations. The vast majority of these
planets have a projected mass, $m\times \sin i$, of about one Jupiter mass ($\mjup
\simeq 10^{-3}\,\msol$) but are found to orbit much closer to their parent star
than Jupiter, with a pile-up around $a\sim 0.04$ AU. As such, these planets
experience strong irradiation from their parent star.  Observations of objects
that transit their parent star provide information crucial to understanding
these so-called hot-jupiters.  The transit light curve or photometry,
supplemented by radial velocity measurements of the star, yields 
the absolute mass and average radius of the planet.  Together with an
estimate of the age of the parent star from stellar evolution models, the
radius and mass provide stringent constraints on the evolution of the
internal and atmospheric properties of strongly irradiated planets.  The
correct description of these properties has been questioned recently by the
detection of the transit planet $\hd$ (Brown et al. 2001). No consistent
model can adequately reproduce the observationally determined mass and radius
of the planet ($R_p=1.42^{+0.10}_{+0.13} \,\rjup$) (Cody \& Sasselov 2002)
without stretching these determinations beyond their error bars (Guillot \&
Showman 2002, Baraffe et al. 2003, Burrows et al. 2003). This led to the
suggestion that an important part of fundamental physics might be missing in
the description of irradiated planets and has motivated several authors to 
propose possible shortcomings of the theory. In the present letter, the
theory developed in Baraffe et al. (2003) for non-irradiated and irradiated
planets is applied to various cases of hot-jupiters, with special focus on the case
of the second observed transit, $\tr$. The theory adequately reproduces the
observed radius of the planet, for the proper age of the system and
the more recent mass determination (Torres et al. 2003).
The comparison of the present predictions for the evolution of
hot-jupiters with forthcoming transit detections will allow the unambiguous
verification or rejection of the present theory.

\section{Effect of irradiation}

For a planet in close orbit around its parent star, the strong incident
stellar radiation substantially modifies not only the emergent spectrum but
also the evolution of the irradiated planet, as shown initially by Guillot et
al. (1996). These effects are examined below.

\subsection {Atmosphere}

Under strong irradiation conditions, the solution of the transfer equation,
which determines the atmospheric thermal structure of the planet, must include
the incoming incident stellar flux in the source function.  Such calculations
were first conducted for hot-jupiters by Seager \& Sasselov (1998), using a
limited set of opacities, and have subsequently been improved by Barman, Allard
\& Hauschildt (2001) and Sudarsky, Burrows \& Hubeny (2003). Figure
\ref{f1.eps} compares the thermal structures of irradiated and
non-irradiated planets with surface gravity $\log g=3.5$ (cgs units) and
intrinsic temperatures $\te=500$ K and $\te=100$ K, respectively, orbiting a G2
main sequence star at 0.023 AU or 0.046 AU. The atmosphere models were produced
using the same cloud-free opacity setup described by Allard et al. (2001) and Barman, Allard \&
Hauschildt (2001).  As examined below, this $\te$ sequence corresponds to an age
sequence of $\sim$3$\times 10^7$ yr to $\sim$9$\times10^{9}$ yr
for $\tr$. The irradiated thermal profile is shown to be
strongly modified with respect to the non-irradiated one.  As stressed in
Baraffe et al.  (2003), the differences between irradiated and non-irradiated
temperature profiles clearly demonstrate the necessity to do {\it consistent}
calculations that combine the irradiated atmosphere profile and interior
profile.  Matching arbitrarily the inner profile to an atmospheric profile
defined by the equilibrium temperature of the planet yields severely flawed
results.  

Note that the present calculations assume that all the incident flux is
concentrated on the day-side of the planet. Such an assumption maximizes the
effect of irradiation and might overestimate the flux deposit. The present
nearly isothermal irradiated profiles, however, are consistent with the
2$\mu$m observed spectrum of $\hd$ (Richardson, Deming \& Seager 2003).  
Future calculations will include the latitude dependence of the incoming
radiation but it is certainly interesting to verify whether the present
calculations, with minimal assumptions, adequately reproduce the radius of new
observed transiting planets.

\subsection{Evolution}

Consistent evolution calculations of irradiated planets have been conducted
only recently by Baraffe et al. (2003) and Burrows et al. (2003). The
calculations presented here proceed as described in Chabrier \& Baraffe (1997)
for the non-irradiated case. A grid of irradiated atmospheric structures is
computed for various $(\log g, \te)$ conditions, for a given orbital distance
$a$, and interpolation between these structures yield the unique boundary
condition with the internal structure, at large enough optical depth ($\tau
\approx 100$).  
As shown initially by Guillot et al. (1996), irradiation pushes
the radiative-convective boundary deeper into the interior and,
under strong irradiation conditions, may force the boundary between the
atmospheric and internal structure to lie in a radiative layer. 
To insure consistency between the atmosphere and internal thermal structures,
we have computed the Rosseland mean of the atmospheric opacities and used this
value to calculate the radiative gradient for the interior structure. The
evolution of the planet obeys the usual first and second principles of
thermodynamics. The energy balance, however, must include the incoming stellar
radiative flux ${\mathcal F}_{inc}={1/2 }({R_\star / a})^2{\mathcal F}_\star$,
where ${\mathcal F}_\star$ and $R_\star$ denote the parent star flux and
radius, respectively.  The total emergent flux of the planet ${\mathcal
F}_{out}$ now reads :

\begin{eqnarray}
{\mathcal F}_{out} &=&{\mathcal F}_{inc}+\sigma \te^4 \nonumber \\
    &=&A\,{\mathcal F}_{inc}\,+\,(1-A)\,{\mathcal F}_{inc}\,+\,\sigma\,\te^4
\label{evolir}
\end{eqnarray} 

\noindent where $\te$ denotes the intrinsic effective temperature of the
planet, $A$ its Bond albedo, and the first term on the r.h.s. of
eqn.(\ref{evolir}) describes the reflected part of the spectrum. Note that the
albedo $A$ is not a free parameter in our calculations but is calculated
consistently with the radiative transfer equation. For both transit cases,
the models predict $A < 0.1$. The total luminosity thus reads:

\begin{eqnarray}
L_{tot} &=&L_{reflected} \,+\,4\pi R^2_p\sigma (T_{eq}^4+\te^4)\nonumber\\
        &=&L_{reflected} + 4\pi R^2_p\sigma T_{eq}^4 - \int T {d S\over dt}dm
\label{evolirrad}
\end{eqnarray} 

\noindent where $S$ denotes the specific entropy of the planet. Whereas the
total flux ${\mathcal F}_{out}$ or luminosity $L_{tot}$ is the quantity
accessible to observation, only the last term on the right hand side of
eqn.(\ref{evolirrad}) concerns the evolution of the planet intrinsic luminosity
$L_{int}$. The first term, which defines the Bond albedo, illustrates the
fraction of the stellar luminosity reflected by the planet atmosphere.
The second term defines its equilibrium temperature, i.e. the
temperature the planet would reach in the absence of any internal source of
energy ($\te \rightarrow 0$).  
Figure \ref{f2.eps} compares the emergent flux calculated for $\tr$ and $\hd$ at
their present ages.  In both cases, the infrared part of the spectrum ($\simgr
1\,\mu$m) is dominated by the re-radiation of the absorbed incident stellar
flux.  Indeed, under the conditions of interest, this contribution largely
dominates the intrinsic contribution of the planet ($T_{eq}\simeq2400$ K
$\gg T_{eff}\simeq$100 K for $\tr$, and $T_{eq}\simeq 1700$ K $\gg T_{eff}\simeq$100 K for $\hd$).  For $\hd$, the short wavelength part of the spectrum
($\simle 0.5\,\mu$m) is mostly due to reflection of stellar light by H$_2$
Rayleigh scattering. However, in $\tr$ the high equilibrium temperature,
$T_{eq}=2400$ K, leads to a lower concentration of H$_2$.  Consequently,
reflection is less significant and the majority of the spectrum of $\tr$ is
thermal radiation.

\begin{figure}
%\centerline{\bf FIGURE CAPTION}

%\begin{figure}
\plotone{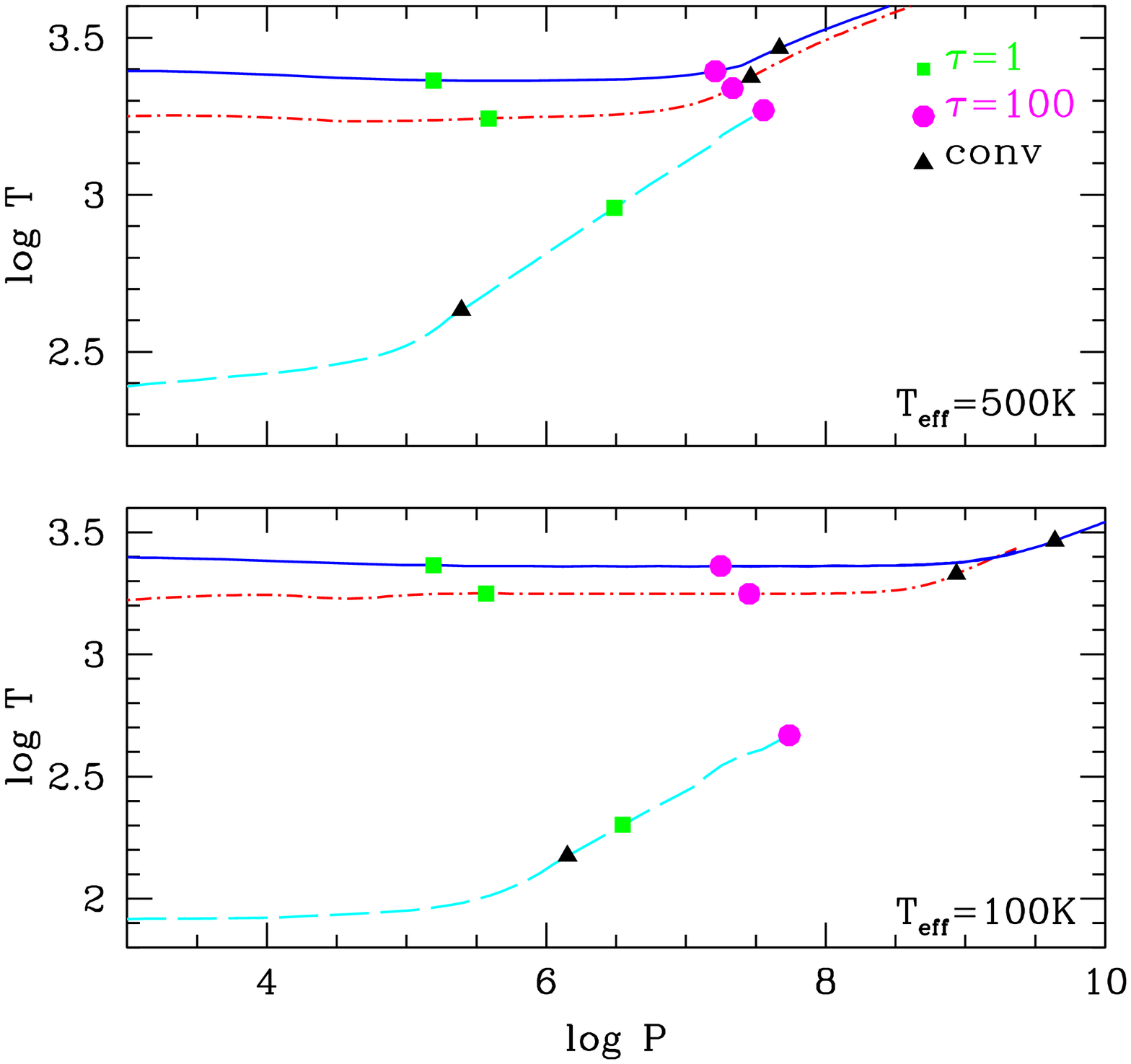}
\caption{Atmospheric profiles (P in cgs units) down to the internal adiabat for non-irradiated
(long-dash line) and irradiated planets with $\te=500$ K and $\te=100$ K, $\log
g=3.5$ orbiting a G2 star at different orbital distances: $a=0.046$ AU
(dash-dot line) and $a=0.023$ AU (solid line).  Symbols indicate various
optical depths (defined at $\lambda = 1.2 \mu m$) and the top of the
convective zone.
\label{f1.eps}}
\end{figure}

\begin{figure}
\plotone{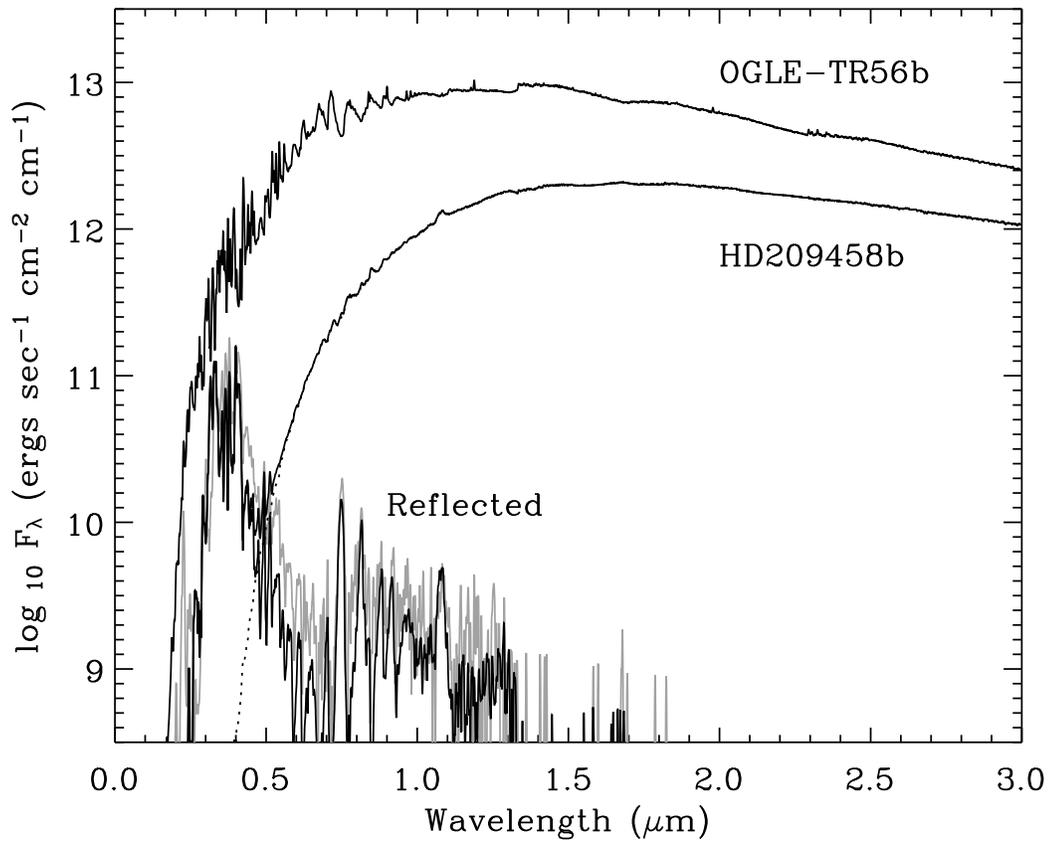}
\caption{
The emergent fluxes for $\te=100$ K irradiated models
corresponding to $\tr$ ($\log g=3.5$) and $\hd$ ($\log g=3.0$).
The lower two curves that peak between 0.3 and 0.4 $\mu$m are the
reflected fluxes for $\tr$ (grey) and $\hd$ (black).  The dotted
line is the continuation of the thermal flux for $\hd$. All
spectra have been reduced to a resolution of 20\AA.
\label{f2.eps}}
\end{figure}

\begin{figure}
\plotone{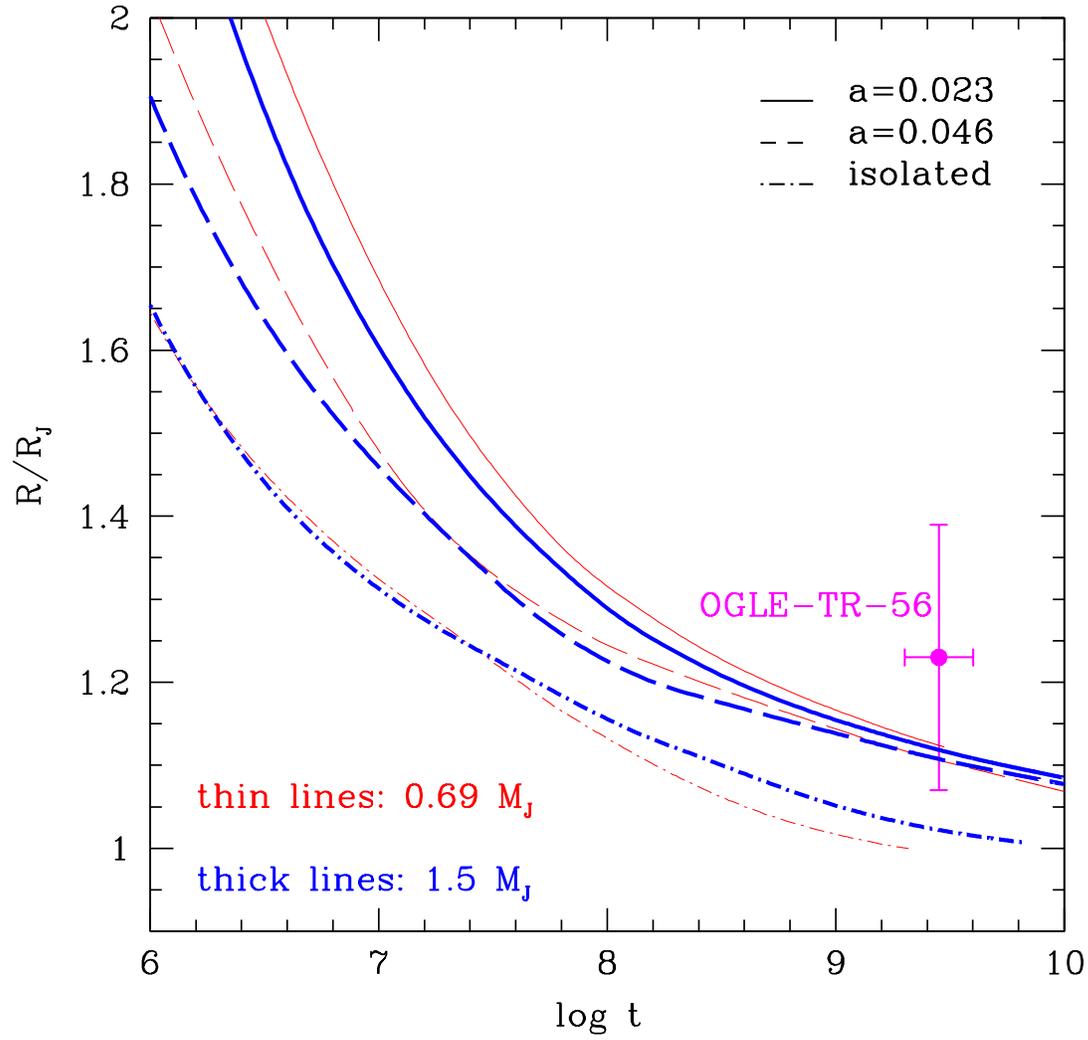}
\caption{Evolution of the radius, in units of Jupiter radii, of irradiated
planets with different mass: m=0.69 $\mjup$ (thin lines) and m=1.5 $\mjup$
(thick lines), orbiting a G2 star at different orbital distances: $a=0.023$ AU
(solid lines) and $a=0.046$ AU (dashed lines).  The case 1.5 $\mjup$, $a$=0.023
corresponds to $\tr$.  Dash-dotted curves display the evolution in the
non-irradiated case, corresponding to planets far away from their parent star
($a \gg 1$ AU).}
\label{f3.eps}
\end{figure}

Figure  \ref{f3.eps} displays the evolution of the radius for different
strongly irradiated planet conditions, with masses 0.69 $\mjup$ and 1.5
$\mjup$, orbiting a G2 star (${\te}=5900$ K) at different orbital
distances, namely $a=$0.046 AU and $a=$0.023 AU. At a given orbital distance, the
less massive the planet the larger the effect of irradiation.  This was
expected from Figure  \ref{f1.eps}, showing the larger modification of
the inner atmosphere profile with larger stellar/planet flux ratio. In all
cases, however, irradiation substantially slows down the contraction of the
planet, in particular at the early stages of evolution, when degeneracy effects
in the interior are smaller.
The case $m=1.5\,\mjup$ and $a=$0.023 AU (thick solid line) corresponds to
$\tr$, for its recently revised mass, $m\simeq 1.45\pm0.23\,\mjup$, and
radius $1.23\pm 0.15\,\rjup$ (Torres et al. 2003).  The age of the system was
derived from the evolution of the parent star for its observed
extinction-corrected magnitude and effective temperature (Sasselov 2003). 
We derive $m\simeq 1.05\,\msol$, $t\sim 4$ Gyr to reproduce the observations, in
excellent agreement with Sasselov (2003).  
The striking result is the very good agreement, within the error bar, between
the present calculations and the observations.  A prediction of the theory is
that irradiation from a G2 star yields less than about $15\%$ increase of the
planet radius after $\sim 1$ Gyr, compared with the non-irradiated case, with
all radii merging within the $\sim 1.1$-1.15 $\rjup$ range at this stage.
A key test for the theory, however, would
be the observation of young $(\simle 10^9$ yr) transits, where radii of
irradiated planets are predicted to vary rapidly with time and to depend
substantially on the orbital distance.

The case of $\hd$ thus becomes a puzzle. As discussed at length
in Baraffe et al. (2003), no consistent evolution calculation can reproduce the
observed radius.  Note that our calculations take into
account the extension of the atmosphere, which we find to be of the order of
$\sim 0.05\,\rjup$ (at optical wavelengths) for the age of this object. Given
the remaining uncertainties in the opacities of these planets, including for
example non-equilibrium processes or day-night temperature gradients, we cannot
exclude larger extensions. It is very unlikely, however, that different
opacities modify the internal adiabat strongly enough to increase the planet
radius by 30\% and thus solve the aforementioned discrepancy between theory and
observation for $\hd$. Such a modification requires a drastic change of the energy
deposited at the top of the internal adiabat, which essentially determines the
radius (Baraffe et al. 2003, Guillot \& Showman 2002). 
Molecular absorption, however, will prevent incident photons from penetrating that
deep ($T\sim 2000$ K near the top of the internal adiabat, see Figure
\ref{f1.eps}). As discussed in \S4.3 of Baraffe et al. (2003), dynamical
processes such as tidal dissipation or synchronization are unlikely to provide
the extra source of energy, for these timescales are at most of the order of
$10^8$ yr.  An interesting alternative scenario involves the inner
redistribution of large scale atmospheric kinetic energy (winds) generated by
the incident irradiation (Guillot \& Showman 2002, Showman \& Guillot
2002).  Besides arbitrarily modifying the atmosphere-interior boundary condition, however,
these calculations face a similar issue as mentioned above, namely
about 1\% of this kinetic energy 
must {\it constantly} be dissipated at the internal adiabat level, i.e.
$P\sim 500$ bar for $\hd$  (see Figure 5 of Baraffe et al. 2003) and $\sim 4\times10^3$ bar
for $\tr$ (see Figure
\ref{f1.eps}). This corresponds to 18 and 23 pressure scale heights for
these two planets, respectively.  Whether such a fraction of the surface flux
of kinetic energy can be transported downward over so many scale heights
and converted into thermal energy remains to be uniquely determined.

\begin{figure}
\plotone{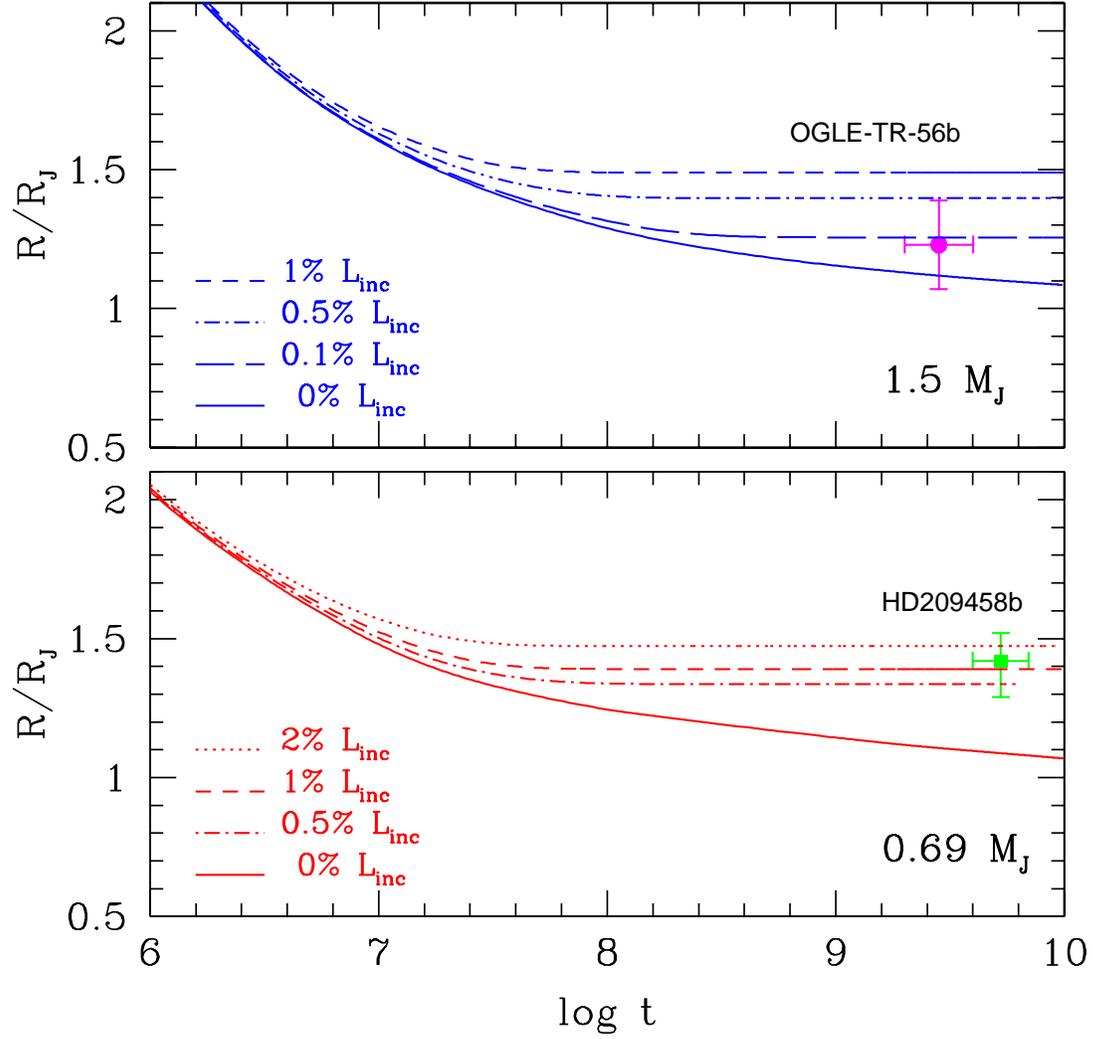}
\caption{Same as Figure \ref{f3.eps} when various fractions of the
incident luminosity $L_{\rm inc}$ are included as an extra source of energy
in the planet interior. The cases are representative of $\tr$ (upper panel)
and $\hd$ (0.69 $M_{\rm J}$, $a$=0.046, lower panel). The 0\% case (solid
lines) correspond to the standard irradiated case. Note that $L_{\rm inc}$ is
$\sim$ 4 times larger in the case of $\tr$  than for $\hd$.}
\label{f4.eps}
\end{figure}

In order to test this scenario in the case of $\tr$, we have made evolution
calculations where a fraction of the incident stellar flux is deposited deep
enough in the planet to modify its internal adiabat (see Baraffe et al. 2003).
Figure \ref{f4.eps} displays the evolution for the two transiting planets
when a fraction $X_\star={\mathcal F}_{dep}/{\mathcal F}_{inc}$ of the incident
flux over the day-side surface is deposited in the interior, with $X_\star
\simeq$ 0.1\% to 1 or 2\%. These values correspond to about 40 to 400 times
the planet's intrinsic energy flux for $\tr$ and up to about 2000 times for  $\hd$.
As shown in the figure, {\it if} indeed
$\tr$ is unambiguously confirmed as a transiting planet, values of $X_\star$
larger than $\sim 0.5\%$ seem to be excluded, for they yield an interior
entropy and radius that are too large.  For $\hd$, on the contrary, $X_\star =
0.5\%$ is about the minimum fraction needed to be consistent with the
observational error bar. As shown earlier, however, the internal adiabat for
the $\hd$ irradiation conditions lies at fewer pressure scale heights from the
surface than for $\tr$ so that we expect intuitively a larger fraction of
energy to be transported downward. 
Although these conclusions must be considered with caution, they suggest that
the dissipative process invoked by Showman \& Guillot may indeed take place in
hot-jupiters.  A tiny ($\simle 0.5 \%$) fraction of the surface kinetic flux
transported downward and converted into thermal energy at the planet's
radiative-convective boundary is enough to modify significantly its internal
adiabat and thus its contraction rate. The result would be a total $\sim
20$-30\% increase of the radius, instead of the aforementioned $\sim 15\%$
increase due to bare insulation.
Another appealing alternative solution to the $\hd$ dilemma is the
suggestion by Bodenheimer, Laughlin \& Lin (2003) of a second orbiting planet,
forcing the finite eccentricity ($e\sim0.03$) of $\hd$. The constant tidal
heating due to the orbit circularization could provide the required mysterious
extra source of energy.

This key issue will be nailed down in the near future with further observations of
transits, as planned with the various ground-based (STARE, VULCAN) and
spaced-based (COROT, MONS, MOST) missions.

\section{Conclusion}

In this letter, we have extended the theory we derived recently for the
evolution of $\hd$ to various conditions of so-called hot-jupiters, i.e.
gaseous planets in close enough orbit around their parent star for the incoming
stellar irradiation to affect substantially their structure and evolution. As
argued previously, a consistent treatment between the atmospheric profile and
the interior profile is mandatory to yield reliable results, both in the
irradiated and non-irradiated case. The theory, in its most simple foundation
(1D irradiation, with incoming flux deposited uniformly over the day side and
no dynamical redistribution of the incident flux due to day-night temperature
differences) is found to adequately reproduce the radius of $\tr$. These
conclusions, of course, rely on $\tr$ being a genuine exoplanet. Recent high resolution
spectroscopic observations seem to support this conclusion (Konacki et al. 2003).
If $\tr$ is confirmed to be, unambiguously, a
planet, then the present results add intriguing
support to the suggestion
that the large radius of $\hd$ might be due to ongoing tidal interaction with
an undetected companion. 
However, we show that a very small fraction of the stellar incident flux,
transformed into large scale kinetic energy partly dissipated at the internal adiabatic
level, is sufficient to modify substantially a planet's internal entropy rate
and thus evolution. This fraction is unlikely to be larger than a few 0.1\%,
but certainly depends on the irradiation conditions.
A fraction $\sim 0.1$ to 0.5 \% would superbly reproduce the observations of
the two presently detected transits, $\tr$ and $\hd$. The present theory allows
us to derive the fundamental properties of hot-jupiters (mass, temperature,
radius) and their observable spectroscopic or photometric signatures as a
function of age and orbital distance. 
Additional transit detections will also allow better constraints on possible irradiation induced
dissipative processes, as explored in this letter.
Comparisons between the present evolution theory and transit or
direct detections of exoplanets will greatly improve our
understanding of these objects and provide strong motivation for
exploring this exciting new field of astronomy.

\acknowledgments

Part of this work was conducted at the Anglo-Australian Observatory, under the auspice of the franco-australian PICS program. This research was supported in part by the LTSA grant NAG 5-3435 to Wichita State University.

\clearpage

\clearpage

\clearpage

\clearpage

\end{document}